\documentclass{acm_proc_article-sp}

\usepackage{xspace}
\usepackage{hyperref}
\usepackage{array}
\usepackage{multirow}
\usepackage{amsmath}

\def\clawpack{Clawpack}
\def\cudaclaw{\textsc{cudaclaw}\xspace}

\newcommand{\Dt}{\Delta t}
\newcommand{\Dx}{\Delta x}
\newcommand{\Dy}{\Delta y}
\newcommand{\imh}{{i-\frac{1}{2}}}
\newcommand{\iph}{{i+\frac{1}{2}}}
\newcommand{\jmh}{{j-\frac{1}{2}}}
\newcommand{\jph}{{j+\frac{1}{2}}}
\newcommand{\bq}{\mathbf q}
\newcommand{\bx}{\mathbf x}
\newcommand{\bu}{\mathbf u}
\newcommand{\bflux}{\mathbf f}
\newcommand{\bglux}{\mathbf g}
\newcommand{\bs}{\mathbf s}
\newcommand{\Aop}{{\mathcal A}}
\newcommand{\Bop}{{\mathcal B}}
\newcommand{\DQ}{\Delta Q}
\newcommand{\amdQ}{\Aop^-\DQ}
\newcommand{\apdQ}{\Aop^+\DQ}
\newcommand{\bmdQ}{\Bop^-\DQ}
\newcommand{\bpdQ}{\Bop^+\DQ}

\begin{document}

\title{CUDACLAW: A high-performance programmable GPU framework for the solution of hyperbolic PDEs}

\numberofauthors{4}
\author{
\alignauthor
H. Gorune Ohannessian\\
       \affaddr{American University of Beirut (AUB)}\\
       \email{gorune@gmail.com}
\alignauthor
George Turkiyyah\\
       \affaddr{American University of Beirut (AUB)}\\
       \email{gt02@aub.edu.lb}
       \and  
\alignauthor
Aron Ahmadia\\
       \affaddr{King Abdullah University of Science and Technology (KAUST)}\\
       \email{aron@ahmadia.net}
\alignauthor
David Ketcheson\\
       \affaddr{King Abdullah University of Science and Technology (KAUST)}\\
       \email{david.ketcheson@kaust.edu.sa}
}

\maketitle

\begin{abstract}

We present \cudaclaw{}, a CUDA-based high performance data-parallel framework for the solution of multidimensional hyperbolic partial differential equation (PDE) systems, equations describing wave motion. \cudaclaw allows computational scientists to solve such systems on GPUs without being burdened by the need to write CUDA code, worry about thread and block details, data layout, and data movement between the different levels of the memory hierarchy. The user defines the set of PDEs to be solved via a CUDA-independent serial Riemann solver and the framework takes care of orchestrating the computations and data transfers to maximize arithmetic throughput.  \cudaclaw treats the different spatial dimensions separately to allow suitable block sizes and dimensions to be used in the different directions, and includes a number of optimizations to minimize access to global memory. 

We demonstrate the power of \cudaclaw{} on 2D and 3D acoustics wave equations and nonlinear shallow water flow simulations.  Analysis shows that even though these simulations are memory bound, sustained performance of more than 180 GFlop/s is obtained on the Tesla C2050 Nvidia GPU, nonetheless. Such performance is comparable to that of manually-tuned code for the mix of floating point operations involved in solving these equations. \cudaclaw includes a high-performance graphics module to view the evolving solution in real time. The \textsc{cudaclaw} framework is accessible through a Python interface, PyClaw. PyClaw was originally developed as an interface to the Clawpack Fortran and PETSc parallel libraries for solving hyperbolic PDEs on distributed memory supercomputers. This very same interface can now also be used for solving these PDEs on GPU-accelerated nodes as well.

\end{abstract}

{Parallel Programming}
{Finite Volume Methods}



\section{Introduction}
We present \cudaclaw{}, a high-performance data-parallel solution framework for 2D and 3D hyperbolic partial differential equation (PDE) systems. Our primary motivation for the development of this framework is to enable computational scientists to solve this broad and important class of simulations efficiently without having to write low-level code or worry about low-level details of data layout and movement between different levels of the memory hierarchy --- details that are essential for obtaining performance on GPUs. Our framework allows scientists to define the PDEs to be solved using a high-level GPU-independent description, and our proof-of-concept results show that the resulting simulations run at speeds comparable to manually tuned code. 

Time-dependent hyperbolic systems of PDEs arise in a broad range of
application domains in engineering and science including acoustics,
elastodynamics, hydrodynamics, and optics.  Computational scientists
and engineers are interested in solving these systems of equations efficiently
and reliably in order to capture complex nonlinear phenomena and understand
shock waves and other characteristics that appear in simulations.
In many cases, the computation of time-sensitive solutions demands higher
performance than what the current generation of workstations offers; for
instance, in the case of tsunami forecasting.  Such forecasts are often
based on solution of the shallow water equations \eqref{eq:SW}, with initial
conditions determined from seismic data and buoys.  In order to be useful,
the simulations must proceed much faster than real-time.

Scientists wishing to perform such numerical simulations on today's manycore
accelerators are faced with a dilemma. On one hand, the promise of
high-performance, inexpensive, multi-teraflop cards offers tantalizing
capabilities for routine use of high-fidelity simulations.  On the other hand,
the gap between the mathematical or algorithmic problem descriptions and the
code that achieves high-performance is wide, making it impractical for
scientists to take advantage of the hardware capabilities. What is needed is a
new generation of systems, such as \cite{fenics,cvx}, that bridge the gap between the expressiveness of
high-level scientific programming languages and the lower-level languages
optimized for execution. 

The \cudaclaw{} framework we describe in this paper is an example of such
software. It is a programmable framework in the sense that it allows a
computational scientist to define a set of PDEs through a ``Riemann solver''
expressed at a high level, shielding the user from the details of data layout,
threads, warps, shared vs. global memory, and the many details that one
normally needs to attend to in order to achieve performance on GPUs. From the
user's point of view, the framework is simple and its use does not require
knowledge of the underlying data structures, nor experience in CUDA
programming. Yet, the code generated is tailored to the GPU architecture,
taking advantage of its arithmetic throughput, and exploiting its high memory
bandwidth and fast on-chip memories. Related efforts have been described in \cite{mint,manyclaw}. 

The solution of hyperbolic PDEs by finite volume methods involves discretizing
the spatial domain of the simulation into cells, with each cell holding a
set of state variables representing solution values. The solution proceeds in time steps, whose size is adaptively
computed from the solution values at the previous time step. At every step, the state
variables in each cell are updated from the values of spatial neighbors. The
primary challenge for obtaining high-performance across the broad spectrum of
hyperbolic PDEs is to abstract the details of data layout and data movement to
shared memory from the arithmetic operations needed to update state variables.
It is this separation that allows the framework to orchestrate data movement
from global to shared memory in an efficient way without user intervention,
and allows the user to specify the arithmetic operations that need to be
performed, independently of how thread blocks operate on the data. A
significant fraction of the computations involved in finite volume methods are
memory-bound. In the roofline model, the maximum performance is achieved on the diagonal
bandwidth-limited portion of the graph. Therefore, GPU code optimizations
generally involve how shared memory is used and how sizes and shapes of 
blocks are chosen to match data access patterns. Optimizations also
involve structuring the computations into kernels that are designed to maximize
their flops-to-byte ratios, thereby improving their performance.

In this paper, we describe the design of \cudaclaw{} and the optimizations it
performs, and demonstrate it on the acoustic wave equation in 2D and
3D, and on the nonlinear shallow-water equations in 2D.  The primary contributions of
the work are a set of GPU-performant algorithms for the solution of hyperbolic
PDEs; an example of a domain-specific scientific
computing system that allows users to customize it with a high level problem
description without sacrificing efficiency;  and a practical system that is
accessible through a Python interface, PyClaw \cite{PyClaw}, to allow scientists to use GPU
acceleration routinely when solving hyperbolic PDEs.  The system is available
from github under \clawpack.  The sample results of our current prototype show that sustained performance of more than 180 GFlops/s is achieved on the Tesla C2050 Nvidia GPU. With the memory-bound
nature of the computations, a roofline analysis shows that this is around 50\% 
of the maximum performance achievable on the C2050, and comparable to the performance of manually-tuned kernels~\cite{Rostrup2010}. 

The rest of this paper is organized as follows. We briefly describe some related prior work in section \ref{prior} and introduce hyperbolic PDEs and the numerical solution that we adopt in section \ref{background_hyperbolic_pde_sec}. In section \ref{cudaclaw_architecture_sec} we give an overview of the structure of the framework, followed by the key design decisions and optimizatios in section  \ref{cudaclaw_system_details_sec}. The Python interface to the system is briefly described in section \ref{pyclaw_integration_sec}.  In section \ref{performance_analysis_sec} we analyze the performance of the framework using the roofline model. Section \ref{future_work_conclusion_sec} concludes. 

\section{Prior Work}
\label{prior}
Because of the importance of hyperbolic PDEs in many engineering and scientific problem domains, substantial work has been conducted, from the early days of GPUs, on developing GPU-friendly algorithms and implementations for their acceleration. 

Shallow water simulations have been performed by a number of researchers.  Hagen et. al. \cite{Hagen2005} implemented a 2nd order central-upwind scheme for 2D shallow water equations with bathymetry and dry states, achieving up to 30x speedup.  In \cite{Lastra2009,Asuncion2010}, first-order schemes for one- and two-layer shallow water flow were implemented in CUDA.  A third-order scheme based on a Roe solver combined with non-oscillatory polynomial reconstruction and Runge-Kutta time integration was implemented in \cite{Gallardo2011}, achieving speedups of more than 100x on a GTX260 card.  This scheme is well-balanced but not capable of handling dry states. Further work by this group, including extension to unstructured triangular meshes, is reported  in \cite{Castro2011}. Brodtkorb et. al. \cite{Brodtkorb2011} implement shallow water schemes that handle dry states as well as friction, and provide direct, realistic visual output at interactive speeds.  They also discuss tradeoffs in using single- or double-precision. Rostrup \cite{Rostrup2010} implement a second-order scheme using a Roe solver, flux limiting, and dimensional splitting, and compare performance on CELL and GPU architectures.

Hagen \cite{Hagen2006,hagen2007solve} describes implementation of first- and second-order  schemes for 2D and 3D flows, with speedups of 10-20x for some standard test problems. Brandvik \cite{brandvik2007acceleration,brandvik2008acceleration} implements 2D and 3D solvers that are 2nd order in space but only first order in time.  Speedups of 15-30x are reported. Kestener et. al. \cite{Kestener2010} implement a first-order Godunov scheme and a second-order central scheme for the 2D Euler equations, achieving 30-70x speedup. Cohen \cite{cohen2010fast} solves the Boussinesq equations with a second-order
scheme and compares GPU acceleration with OpenMP acceleration.  While most of these papers focus on standard test problems, Elsen \cite{ELSEN2008} implements a finite-difference scheme with multigrid for 3D unsteady RANS simulations to solve an industrial problem.

Our work differs from the above in several respects. Rather than focusing on a single set of equations and optimizing the code to take advantage of its particular structure, we are interested in building a framework to handle a wide variety of hyperbolic PDEs, which can vary significantly in the number of variables per cell and in the arithmetic intensity of their core computation. This requires the development of different optimizations for the various stages of the computations as described later. In addition, and in contrast to the usual stencil computations that access the neighboring cells in all spatial dimensions simultaneously in order to compute cell updates, we use dimensional splitting as mentioned in section \ref{background_hyperbolic_pde_sec}. Dimensional splitting allows the PDEs to be solved independently in each dimension, with the solutions of each dimension assembled together to give the final cell update. This strategy significantly enhances the overall memory and cache behavior as it allows data access in the separate spatial dimensions to be optimized more effectively.

\section{Hyperbolic PDEs and Clawpack}
\label{background_hyperbolic_pde_sec}
    Hyperbolic PDEs arise from modelling in various disciplines such as
    engineering, medicine, geology, etc. Hyperbolic PDEs describe
    wave motion. 
    The numerical methods in \cudaclaw{} compute approximate solutions of
    systems of hyperbolic conservation laws which we describe, for simplicity, in two dimensions:
    \begin{align} \label{eq:conslaw}
      \bq_t + \bflux(\bq)_x + \bglux(\bq)_y & = 0.
    \end{align}
    Here $\bq(\bx,t)$ is a vector of conserved quantities (e.g., density,
    momentum, energy) and $\bflux,\bglux$ represent the flux components.
    Here we describe high-resolution shock capturing methods, which
    are one of the most successful classes of numerical methods for solving \eqref{eq:conslaw}.

    Computing solutions to nonlinear hyperbolic equations is often costly.
    Solutions of \eqref{eq:conslaw} generically develop
    singularities (shocks) in finite time, even if the initial data are
    smooth. Accurate modeling of solutions with shocks or strong convective character requires computationally
    expensive techniques, such as Riemann solvers and nonlinear limiters.

    In a finite volume method, the unknowns at time level $t^n$ are taken to be the averages of $q$
    over each cell:
    \begin{align}
        Q^n_{i,j} = \frac{1}{\Dx\Dy} \int_{y_\jmh}^{y_\jph}\int_{x_\imh}^{x_\iph} q(x,t^n) \, dx,
    \end{align}
    where $(\Dx,\Dy)$ and $(i,j)$ are the local grid spacing and the
    cell index, respectively.

    The classic Clawpack algorithm is based on the second-order Lax-Wendroff difference scheme
    that was later extended by LeVeque \cite{leveque1997,levequefvmbook}.
    In two dimensions, it takes the form
    \begin{align}
    \begin{aligned} \label{LW-update}
Q_{i,j}^{*}  = Q_{i,j}^{n} &
     -\frac{\Dt}{\Dx}\left(\apdQ_{i-1/2,j}+\amdQ_{i+1/2,j}\right) \\
    &-\frac{\Dt}{\Dx}\left(\tilde{F}_{i+1/2,j}-\tilde{F}_{i-1/2,j}\right) \\
Q_{i,j}^{n+1} = Q_{i,j}^{*} &
    -\frac{\Dt}{\Dy}\left(\bpdQ^*_{i,j-1/2}+\bmdQ^*_{i,j+1/2}\right) \\
    &-\frac{\Dt}{\Dy}\left(\tilde{G^*}_{i,j+1/2}-\tilde{G^*}_{i,j-1/2}\right).
    \end{aligned}
    \end{align}
    Equation \eqref{LW-update} is a Godunov split scheme, second-order update of the cell average state $Q_{i,j}$
    from time $t^n$ to an intermediate state and then to $t^{n+1}$. The first two terms represent the effect
    of fluxes on the horizontal direction (approximating the term $\bflux(\bq)_x$)
    while the third and fourth terms represent vertical fluxes (the term $\bglux(\bq)_y$).
    The latter two terms are computed using the intermediate cell states after the first two terms are applied.
    The first and third update terms give a first-order update, while the
    second and fourth terms represent a second-order correction that is computed
    using a nonlinear {\em wave limiter}.

    All of the terms in \eqref{LW-update} are computed by solving Riemann problems.
    A Riemann problem consists of a hyperbolic PDE \eqref{eq:conslaw} together with
    piecewise constant initial data composed of two states with a single discontinuity
    between them.  Conceptually, one may think of the finite volume solution as being
    constant within each grid cell; then at each cell edge the local solution corresponds
    to a Riemann problem.  The method may be thought of as solving these Riemann problems,
    evolving the solution over a short time $\Dt$, and re-averaging the solution over
    each grid cell.  The most expensive step is the solution of the Riemann problems.
    These Riemann problems are independent from one another, and provide the
    key to the parallelism that we exploit with the GPU architecture.
    The wave limiter is also a relatively expensive part of the computation; it involves
    taking the waves computed by the Riemann solver and modifying them to add dissipation
    in the vicinity of a discontinuity.

    The time step size $\Dt$ used in \eqref{LW-update} must be chosen carefully.
    Typically, one wishes to take it as large as possible for computational
    efficiency, but numerical stability requires that the step size satisfy
    \begin{align} \label{CFL-condition}
    \Dt & \le C\frac{\Dx}{s}
    \end{align}
    where $s$ is the magnitude of the fastest wave speed occurring in the problem
    and $C$ is a constant depending on the numerical method.  The restriction
    \eqref{CFL-condition} is referred to as a {\em CFL condition}.


    \clawpack{} is a very general tool in the sense that it is easily adapted
    to solve any hyperbolic system of conservation laws.
    The only specialized code required in order to solve a
    particular hyperbolic system is the Riemann solver routine.  A wide range of
    Riemann solvers, including several for the most widely studied hyperbolic systems, have
    been developed by \clawpack{}  users and are also freely available.
    Non-hyperbolic source terms ($\bs(\bq,\bx)$) can be easily included via operator splitting.
    For more examples and details regarding Clawpack,
    see \cite{leveque1997} and \cite[Chapter 23]{levequefvmbook}.

To illustrate the use of our tool, we solve the acoustic wave equation 
\begin{subequations} \label{eq:acoustics}
\begin{align}
p_t + \nabla \cdot \bu & = 0 \\
\bu_t + \nabla p & = 0
\end{align}
\end{subequations}
in 2D and 3D (here $p,u$ are the pressure and velocity), and the two-dimensional shallow water equations 
\begin{subequations} \label{eq:SW}
\begin{align}
h_t + (hu)_x + (hv)_y & = 0 \\
(hu)_t + \left(hu^2 + \frac{1}{2}gh^2\right)_x + (huv)_y & = 0 \\
(hv)_t + (huv)_x + \left(hv^2 + \frac{1}{2}gh^2\right)_y & = 0.
\end{align}
\end{subequations}
Here $h$ denotes the water height while $u$ and $v$ denote the $x$- and $y$-component of the
velocity, respectively.

\section{CUDACLAW Architecture}
\label{cudaclaw_architecture_sec}

  \begin{figure*}[t]
 \begin{center}
 	\scalebox{0.75}{\includegraphics{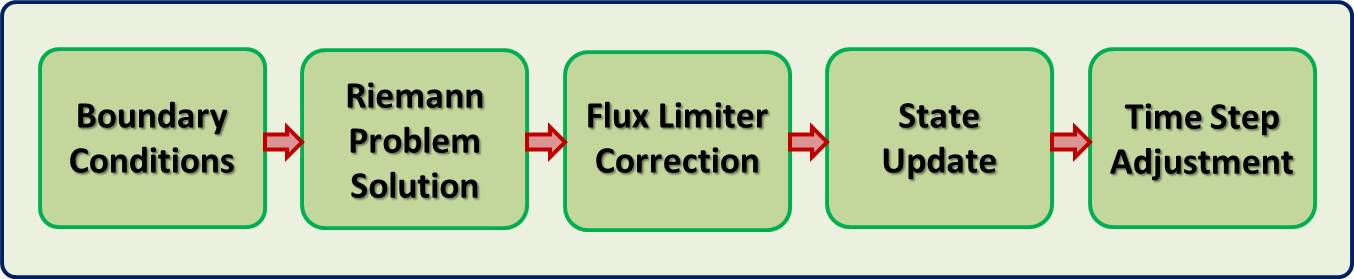}}
 \end{center}
  \caption{CUDACLAW conceptual pipeline \label{abstract_pipeline_fig}}
 \end{figure*}

Based on \clawpack, \cudaclaw{} aims at exploiting the independence of the Riemann problems at cell interfaces and the split dimensional updates to achieve performance on GPUs. Conceptually, \cudaclaw{} implements the pipeline appearing in figure \ref{abstract_pipeline_fig}. It orchestrates data movements between global and shared GPU memory and rearranges computations to eliminate otherwise unavoidable expensive synchronization. It does this by combining stages of this pipeline into kernels ---with relatively high flop-to-byte ratio and adapted memory access pattern--- for efficient execution. \cudaclaw{} also minimizes the memory footprint of the computation, by computing the full wave details in shared memory and avoiding their storage in global memory.  In this section, we describe the overall computations involved. We elaborate on some of the design decisions in the following section.

\subsection{CUDACLAW Computational Pipeline}
As shown in Figure \ref{abstract_pipeline_fig}, our framework is composed of several stages: Boundary Conditions, Riemann Problem Solution, Flux Limiter Computation, State Update and Time Step Adjustment. This computational pipeline grants great flexibility to the framework by allowing boundary conditions, Riemann solvers and limiters to be swapped, furthermore it automatically handles second-order computations leaving the user to define only their Riemann solver and optionally additional limiter functions and boundary conditions.

The framework implements this pipeline as a set of GPU device kernels. The first applies the boundary conditions. The second combines all inner stages, Riemann solution, flux limiting, second-order corrections and state update, there are two(three) of these kernels for the 2D(3D) solver, one for each dimension, and from this point on we will refer to these kernels as `core kernels', in the following section we explain the reason for this split. The third kernel is an auxiliary kernel that reduces over the horizontal and vertical wave speeds, to find the largest absolute speed. The fourth kernel computes the CFL number and decides if the time step should be reverted or the computation continued with a new time step according to equation \ref{CFL-condition}.

In addition to updating the cells inside the domain, the solver must apply
prescribed boundary conditions at the edge of the domain.
Updating the boundaries is independent of the Riemann problems and is
therefore a separate stage and a separate kernel. Usually, boundary conditions are not computationally intensive
and since the boundaries of $2(3)$-dimensional spaces are $1(2)$-dimensional
respectively, this stage is computationally relatively inexpensive.


Once the Riemann problems at cell interfaces are solved, the framework will have enough data to
compute wave limiters, use them in the second-order corrections (lines 2,4 in \eqref{LW-update}) and finalizing the second-order update terms by adding them to the first order update terms (lines 1,3 in \eqref{LW-update}) using a limiter function. The full update terms are then applied to the cells. As mentioned earlier, \cudaclaw{} solves multidimensional problems by dimensional splitting~\ref{LW-update}, solving one dimensional problems along each direction. This makes the inner stages of the pipeline ---Riemann solution, limiter and second order computation--- multistage, where they are repeated in each dimension in the corresponding dimension's core kernel. 

With the finite volume scheme, the Courant, Friedrichs and Lewy (CFL) condition \ref{CFL-condition} states that Riemann solution produced waves must not travel farther than the cells which generated them. This can be ensured by taking a small enough time step, which limits the travel distance of the fastest wave to one cell. This requires the framework to know the fastest wave that was generated, which, on a parallel machine, is done with a reduction operation. We cannot determine such a step before the wave speeds are known, therefore an estimate is used based on the wave speeds of the previous time step and used in the current step. If the CFL was found to be violated the step is reverted and the computation redone with a more appropriate time step. The fastest wave speed can be obtained efficiently on a GPU by a reduction operation over all the wave speeds. With a large number of waves, the reduction can be time consuming. In addition, storing wave speeds in global memory limits the maximum simulation size we can run on the GPU. In \cudaclaw, waves and wave speeds are generated and stored in shared memory (Figure \ref{solver_mem_interaction_fig}) where local reduction on the wave speeds is performed 
and only their local maximum written to global memory. This greatly reduces the number of elements to reduce in global memory.

\subsection {Point-wise Riemann Solver}
\begin{figure}[ht]
 \begin{center}
 	\scalebox{0.33}{\includegraphics{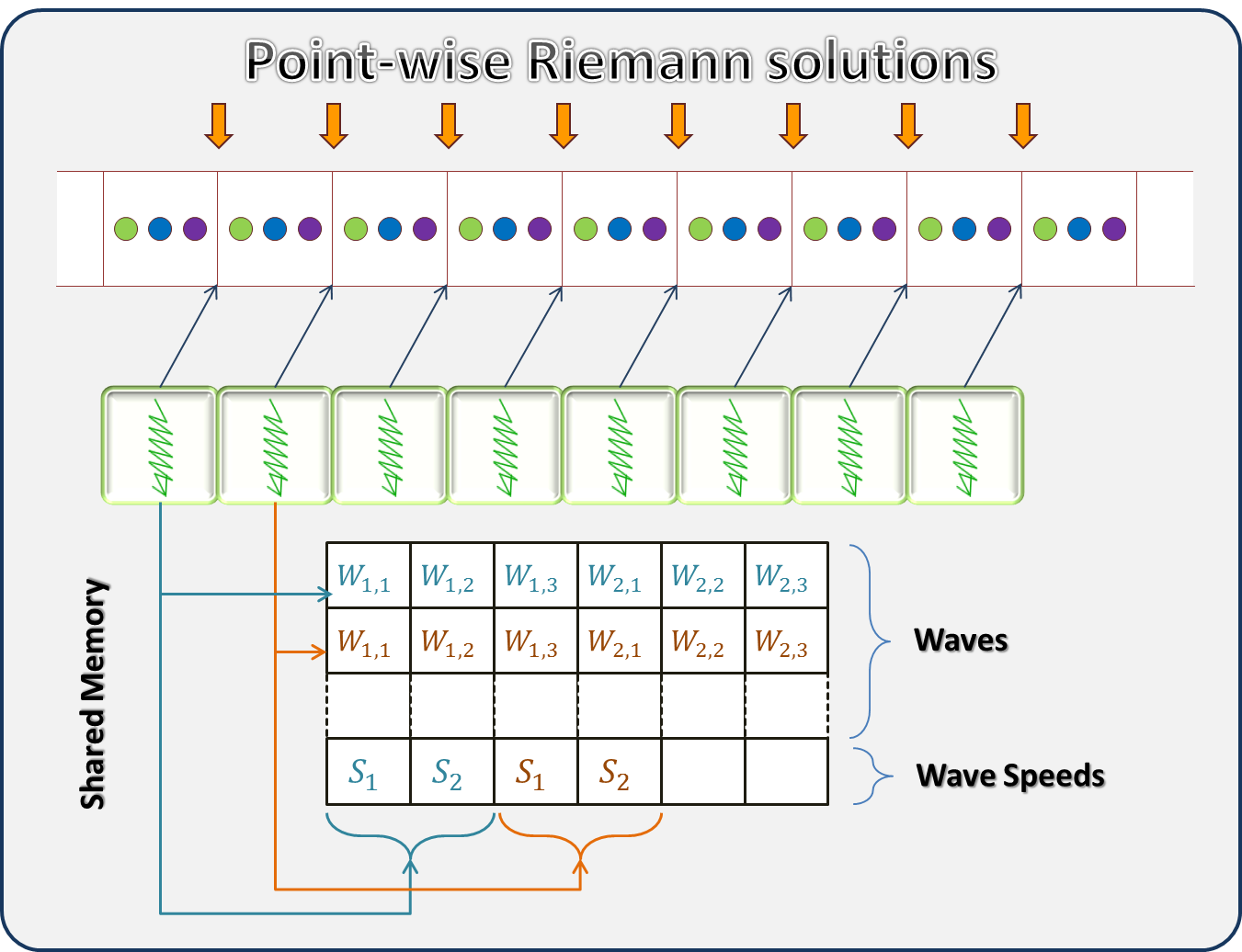}}
 \end{center}
  \caption{A block of threads solving Riemann problems at individual interfaces, using point-wise Riemann solvers. Results are stored in shared memory\label{point_wise_Riemann_fig}}
 \end{figure}

Unlike \clawpack's row-vectorized Riemann solver, we use point-wise Riemann solvers defined as functions that operate on two neighboring cells to generate waves governed by the PDEs being solved.  Riemann solvers are more naturally formulated as scalar functions, but are often coded as vectorized functions to improve computational performance.  However, and as shown in~\cite{manyclaw}, a well-designed computational framework allows application scientists to achieve the performance of architecture-tuned vectorized functions while enjoying the simplicity of working with scalar Riemann kernels that are independent of underlying memory structure and specific indexing arithmetic.  Point-wise Riemann solvers are natural to the GPU's thread level parallelism. Figure \ref{point_wise_Riemann_fig} shows how threads can be mapped to solve Riemann problems at cell interfaces, essentially making the block of threads a vector solver. This also allows for further optimizations in terms of memory savings as point-wise functions are not required to store the first-order update terms in global memory. These are computed using the generated waves and wave speeds available in local memory and immediately used for the second-order correction and update operations.

\subsection{GPU Technical Considerations}
\begin{figure}[ht]
 \begin{center}
 	\scalebox{0.44}{\includegraphics{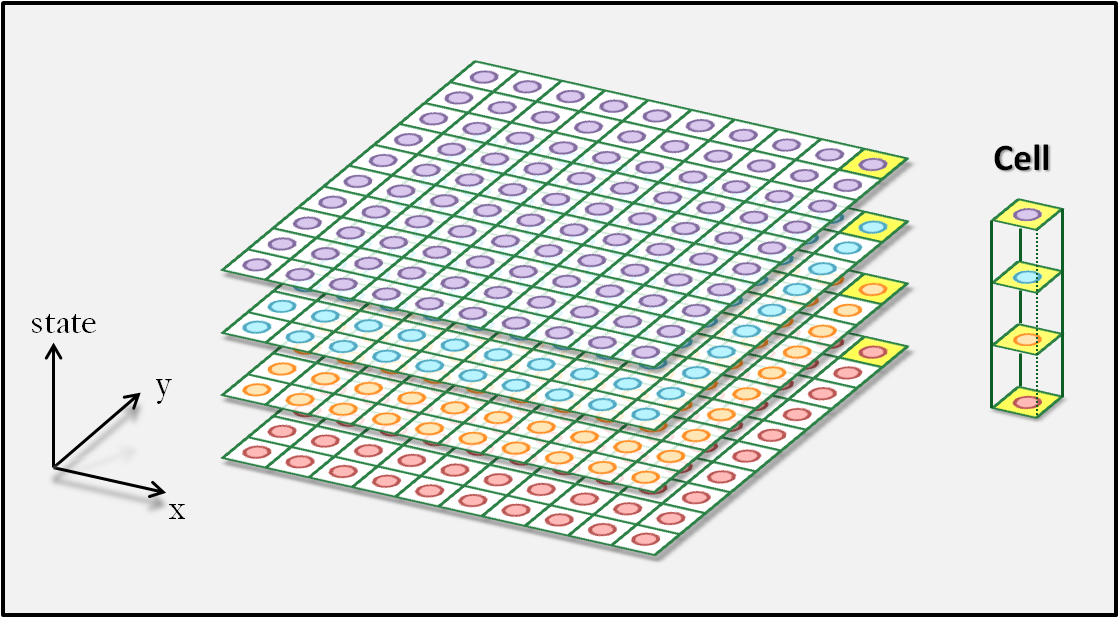}}
 \end{center}
  \caption{CUDACLAW 2D data grid layout in GPU memory: each state is stored contiguously in row major order\label{data_distribution_fig}}
 \end{figure}

The memory layout of this grid is one of the important factors in maximizing memory bandwidth throughput, and as a consequence, computational throughput.
As the finest grain of parallelism of the GPU, the threads of a warp all execute the same instructions, and therefore request memory at the same time. The hardware can satisfy all memory requests within a warp most efficiently if the considered memory addresses are on a contiguous piece of memory, and start at an aligned address, resulting in a \emph{coalesced} access. This characteristic of the GPU consequently makes structures of arrays preferable over arrays of structures. With a multidimensional grid of multi state cells, where the same state is accessed simultaneously across the grid, the best memory layout for a 2D/3D problem is to store each state as a separate 2D/3D grid, in row major order, resulting in the layout depicted in figure \ref{data_distribution_fig}.

Another factor that determines the activity distribution of the GPU is thread work assignment. Threads can be assigned to compute the necessary elements to update a single cell, or can compute the Riemann solution at inter-cell interfaces and then cooperate to compute the update terms in \eqref{LW-update}. Although the former scheme of a one-to-one map between threads and cells offers reduced synchronization and inter-thread communication, the requisite large amount of registers per thread makes it infeasible. In this implementation, we opt to map every thread to an interface, where each thread solves the Riemann problem at an interface, and eventually updates a single cell. The details are discussed in the next section, where we describe how  blocks and threads within them are mapped to the computational grid, and the inner workings of the core kernels, specifically how the output of the the Riemann solver is stored and used in second-order computations and update.

\section{CUDACLAW System Details}
\label{cudaclaw_system_details_sec}

 \begin{figure}[t]
\begin{center}
	\scalebox{0.35}{\includegraphics{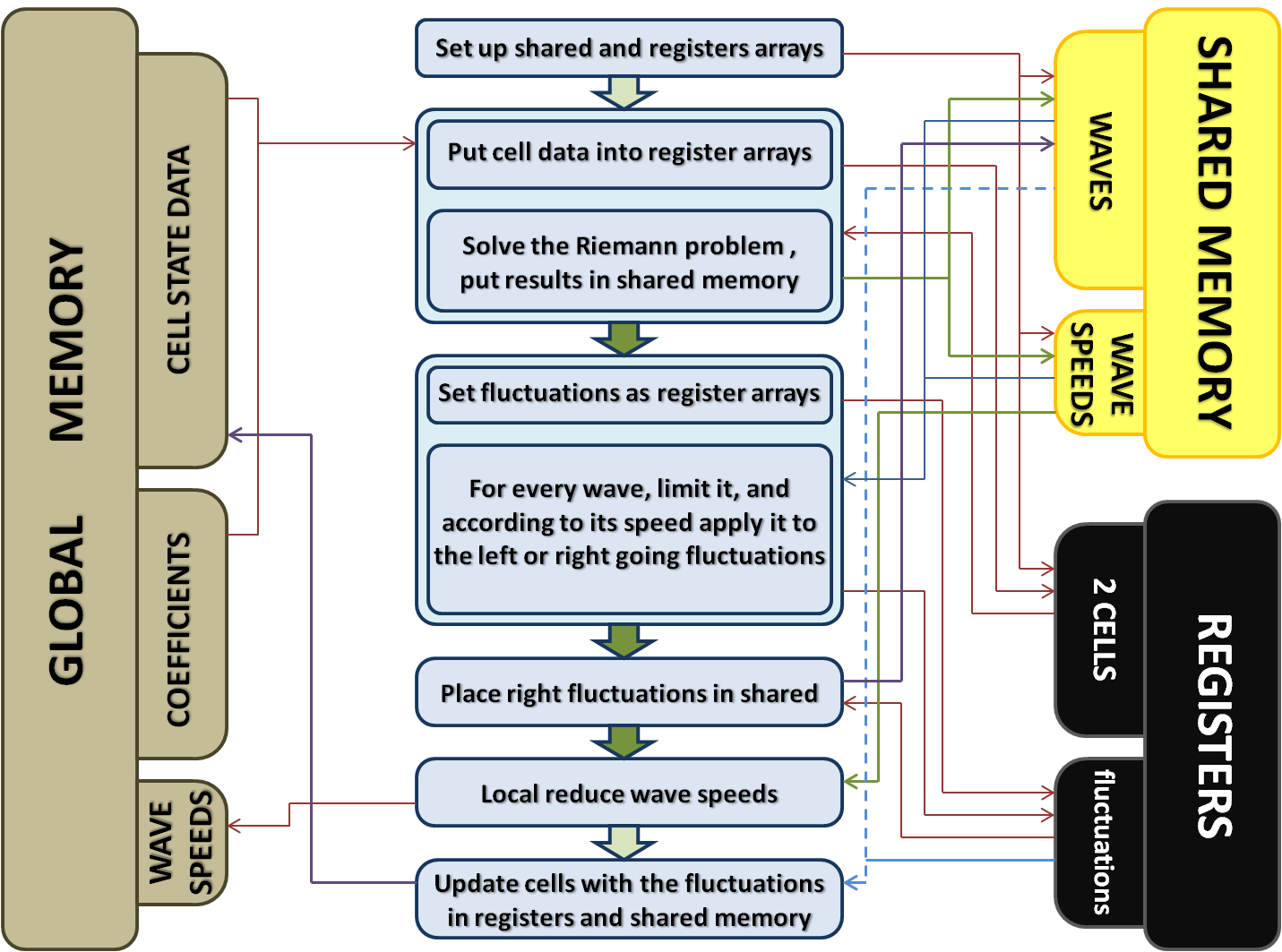}}
\end{center}
 \caption{CUDACLAW's communication patterns with global and fast memories; filled green arrows represent block synchronization} 
\label{solver_mem_interaction_fig}
\end{figure}

In designing the core kernels, we targeted a minimum global memory footprint at the cost of redundant computations. Performing a fractional amount of redundant computations in local memories outweighs the communication and synchronization that would otherwise be required.  Figure \ref{solver_mem_interaction_fig} shows how the stages of a core kernel are implemented, clearly showing the heavy use and reuse of the fast memories, shared memory and registers. Each stage indicated in the figure depends on the output of the previous stages, allowing us to keep data local and avoid global memory. We now fully dissect the core kernel emphasizing the ideas that made this structure possible. 


\subsection {Independent Blocks}
\begin{figure}[t]
 \begin{center}
 	\scalebox{0.35}{\includegraphics{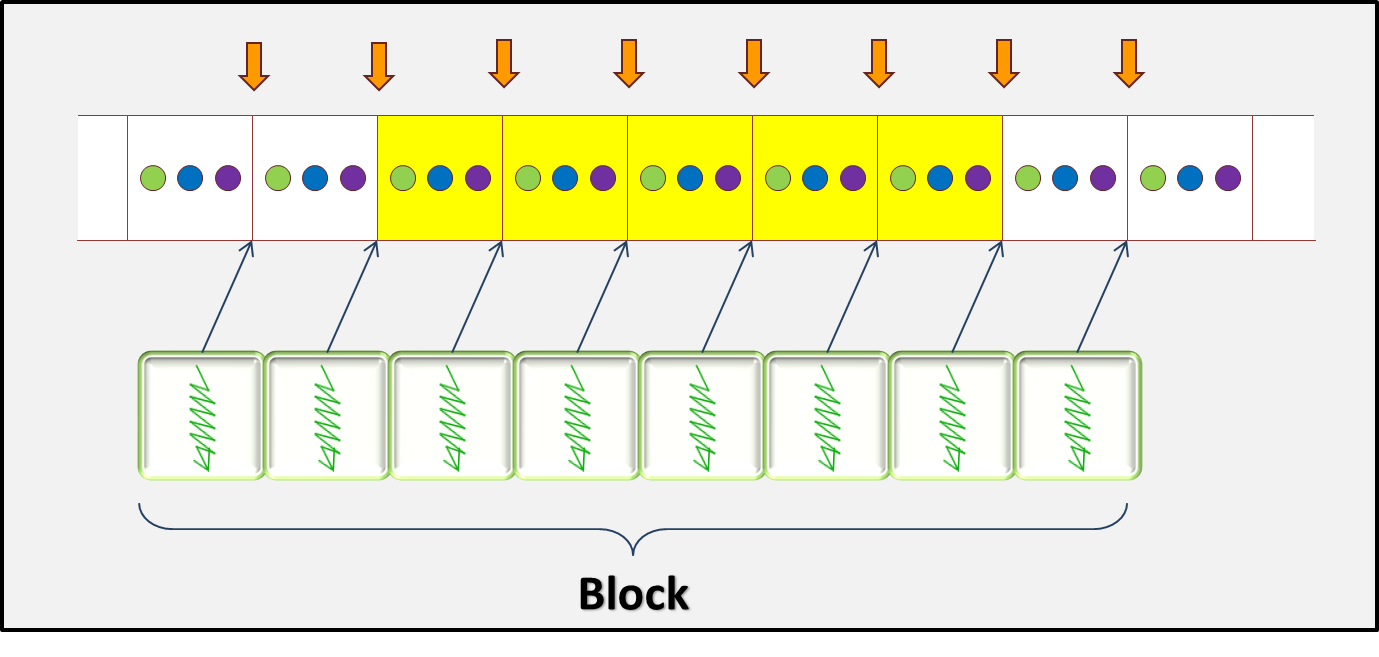}}
 \end{center}
  \caption{A single block is mapped to the highlighted subdomain, threads solve the Riemann problem on interfaces that participate in the update of the subdomain \label{thread_map_fig}}
 \end{figure}

One of the key ideas that allow us to take full advantage of the GPU is block
independence. Instead of mapping threads one-to-one to interfaces, we divide the
computational grid into cell subdomains, and assign each block to update a single
subdomain. The block is then allowed to do all computations needed to
advance the cells in its subdomain. This makes the block completely independent from
other blocks. Figure \ref{thread_map_fig} shows a block assigned to a subdomain.
With a second-order scheme, cell data depends on 5 cells or on 4 sets of waves
from the left and right surrounding interfaces.  In the figure,
required wave information for the highlighted subdomain is indicated by orange
arrows, precisely the interfaces the threads operate on. As the block is
solving an independent subproblem, one can view it as being a full grid, where the equivalent
global memory is the shared memory. As shown in Figure \ref{point_wise_Riemann_fig}, Riemann solutions can be stored
in shared and used later in the kernel, without having to access global memory. Once the
Riemann problem is solved, the border threads will idle, while the rest proceed
to limiter and second-order correction term computation, as these threads do not have one of the necessary two neighbouring wave information in the shared memory.

\begin{figure}[t]
 \begin{center}
 	\scalebox{0.35}{\includegraphics{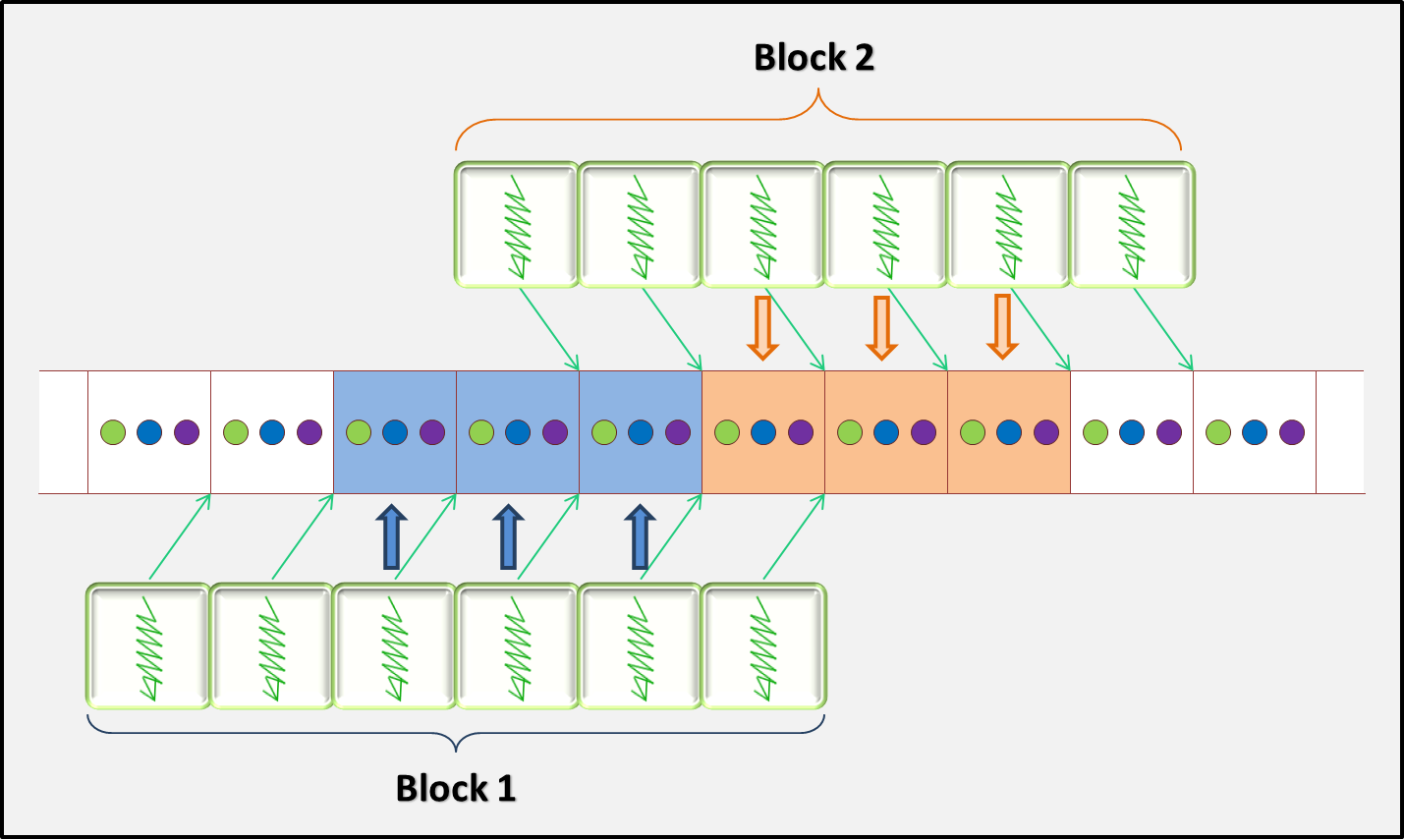}}
 \end{center}
  \caption{Computational and data overlap of two blocks with adjacent subdomains\label{thread_map2_fig}}
 \end{figure}

Such a block map will however incur redundant computations at the interfaces of adjacent subdomains, as shown in Figure \ref{thread_map2_fig}, however we can minimize such redundancy by careful choices of block shape and size (Figure \ref{block_overlap_fig}), and get better performance than we would without any redundant computations. In fact, having independent blocks frees the framework from handling inter-block communication to ensure correctness, which can only happen at the global memory level, which would mean that Riemann solutions must be stored in global memory without independent blocks. Storing the wave structure not only reduces the GPU's capability of solving large problems, it also reduces its throughput with more memory reads and additional expensive synchronization, later discussed in the analysis section.

Given the wave speeds in shared memory, the block can do a local reduction, and write out a single local maximum speed back to global. This reduces the final size of the global reduction by a factor equal to the size of the block times the number of states the problem has.

\subsection {Split Kernel Dimension Computation}
Solving both horizontal and vertical Riemann problems in a unified kernel allows the kernel to read less data, however, on the GPU, such a kernel requires too many registers per thread and shared memory per block as we intend to store the waves and their speeds in shared memory. Also, blocks launched for a unified kernel will have to overlap over data with other blocks from all sides, greatly reducing the number of active threads in the later stages of the solution, namely, limiter and second-order computations. Furthermore it disallows certain optimizations that can be done with kernels dedicated to only one dimension. These drawbacks make such kernel impractical, therefore we use split the computations over the core kernels where each would have its own optimizations dealing with memory transactions and computations of a given dimension. 

\begin{figure}[t]
 \begin{center}
 	\scalebox{0.33}{\includegraphics{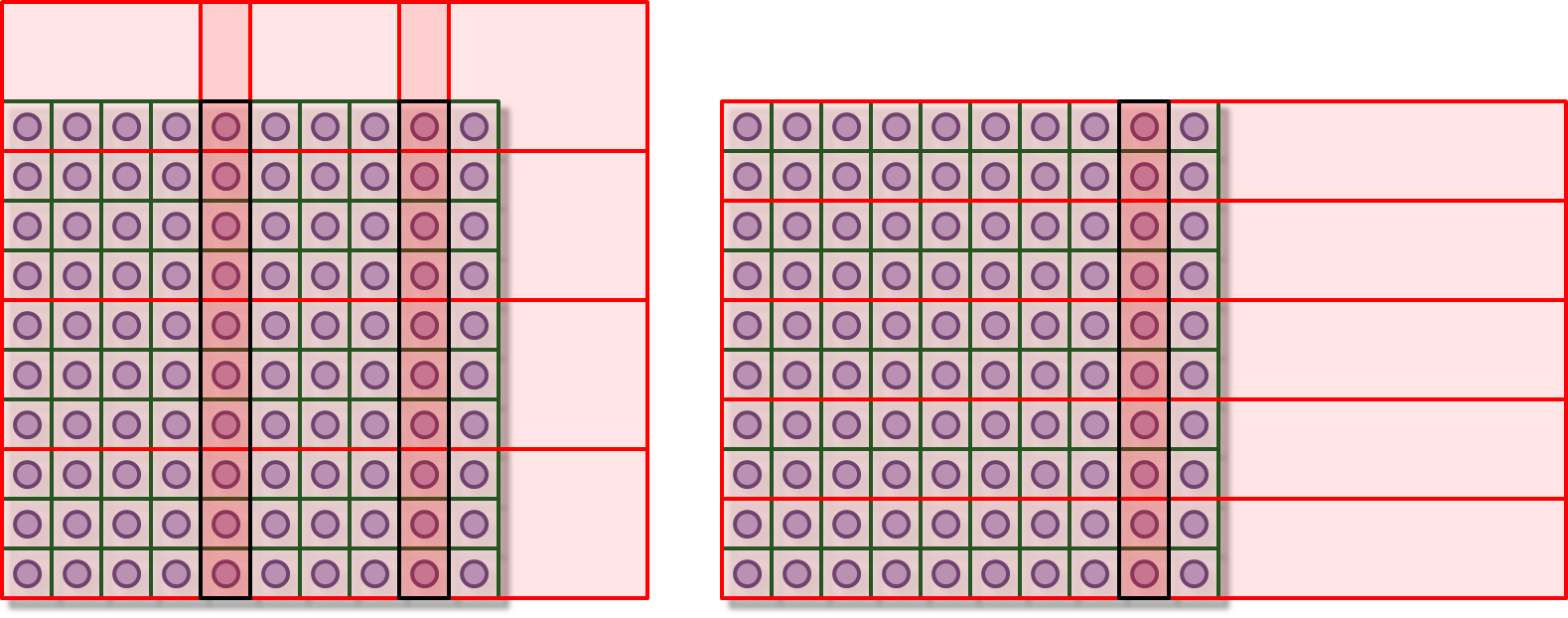}}
 \end{center}
  \caption{In this simplified example, a $3\times 5$ block gives twice as much overlap as a $2\times 9$ block, for the horizontal core kernel\label{block_overlap_fig}}
 \end{figure}

As a kernel is dedicated to either horizontal or vertical Riemann problems, we can choose their shape in order to minimize the block overlap we get by having independent blocks. Note that we use 2D blocks to launch the core kernels of any dimension, for 2D and 3D problems, even though computations are done over a single dimension. Blocks can therefore be viewed as parallel row-vectorized solvers. In terms of memory access patterns, the states in the grid being stored in row major order, computations in the horizontal dimension require overlap over the fastest moving index as shown similar to the blocks Figure \ref{block_overlap_fig}. This makes blocks for the horizontal core kernel start at misaligned addresses, however such misaligned is unavoidable in such stencil computations and will happen in exactly one dimension, while the other kernels get perfect alignment therefore perfect memory transaction per request ratio.

Figure \ref{block_overlap_fig} shows how a $3\times 5$ block gives twice as much overlap as a $2\times 9$ block. A careful tuning is required to get the best performance, as the computational redundancy is not the only factor at play. We found that having a block width such that the length times the data type size that corresponds to at least a complete cache line size yields the best results.

\subsection {Single Stage Update}
For a thread to update a cell in a single stage, update terms from its two interfaces must be available to it. A thread having computed Riemann solutions at a single interface, and used neighbouring wave information to compute wave limiters, will have two second-order update terms in its registers, one for each cell of the corresponding interface. A thread would need the update information from its neighbour's registers, its left neighbour in the horizontal kernel and its bottom in the vertical. However, thread register information cannot be read by any other thread, so the particular information has to be passed from the neighbour to the thread. The closest memory space where such a transaction can happen is the shared memory. By this stage all threads in the block should be done with their previous computations, ensured by a block synchronization, and therefore all wave data is available for re-write. Threads use their space of the first Riemann generated wave in shared memory to store the update term required by their right (top) neighbour. A thread will then be able to update its cell having the proper update terms, one in its registers and one in shared memory.

By contrast, applying the update terms one at a time requires twice as many reading and writing to global memory. This decreases the flop-to-byte ratio of the computation, and reduces performance.



\subsection{Display}
\begin{figure}[t]
 \begin{center}
 	\scalebox{0.5}{\includegraphics{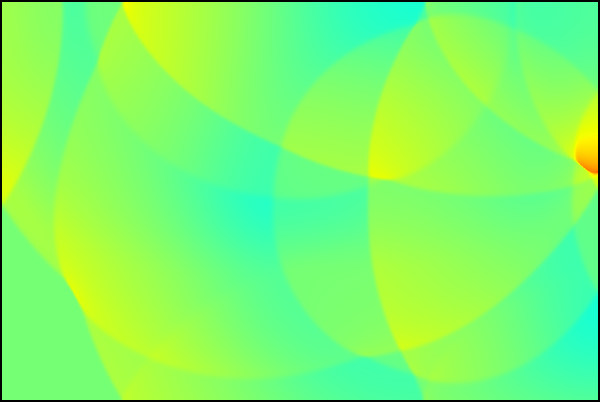}}
 \end{center}
  \caption{A sample shallow water simulation captured as the framework runs\label{shallow_water_example_fig}}
 \end{figure}

As the computational grid remains and is operated on the GPU, \cudaclaw{} offers the option of viewing the solution of the defined hyperbolic PDE at interactive speeds. Given a time adaptive context, the display only outputs frames that reach a given time stamp, resulting in smoothly displayed simulation progression. Figure \ref{shallow_water_example_fig} is a snapshot taken from a shallow water simulation, with reflective boundaries at three interfaces and a transmissive boundary on the bottom. High-performance graphics is possible via the OpenGL-CUDA interoperability Nvidia libraries, making the computational buffer available to OpenGL pixel buffer objects (PBOs).

\section{PyClaw Integration}
\label{pyclaw_integration_sec}

An important aspect of this work is integration into the PyClaw software framework. PyClaw is a Pythonic implementation of the Clawpack algorithms, with performant coupling into the original Fortran Riemann solver routines and a focus on accessibility, extensibility, and scalability. \cudaclaw{} complements several other powerful extensions to the original Clawpack algorithm in PyClaw: SharpClaw and PyWeno provide auto-generated high-order weighted essentially non-oscillatory wave progagation, PetClaw provides performant distributed-computing that scales to tens of thousands of processes, and PeanoClaw is a prototype extension that adds adaptive mesh refinement through the Peano framework.  

Integration with an existing interface provides many advantages. First, we greatly increase the accessibility of our code, and make it more readily deployable. For performance reasons, \cudaclaw{} is necessarily implemented in CUDA C/C++, which require expert knowledge to modify and use to set up customized simulations. PyClaw, on the other hand, is implemented in Python and provides a Pythonic interface.  Python is a language designed to be friendly to beginners without sacrificing performance or expressiveness.  Python scripts are easier to read and customize and would allow more users to have access to \cudaclaw{} In addition, PyClaw has an established user base, who can now take advantage of \cudaclaw{} without learning a new system. 
PyClaw also features an extensive test and application suite, including the acoustics and shallow water examples described in this paper, that users can rely on. Finally, the PyClaw interface allows users to  switch seamlessly between platforms as their needs and access to hardware change. 

A documented prototype PyClaw/\cudaclaw{} interface is one of the software artifacts of this work. The prototype is available at $\clawpack/pyclaw/cudaclaw\_pull\_request$. The prototype features the ability to set up a problem in PyClaw, including the specification of initial and boundary value conditions, then evolve the problem using \cudaclaw's time-steppping algorithm, and verify the computed solution at any subsequent time step.  There are several limitations of the current interface prototype which will need to be addressed.  Most notably, PyClaw currently does not support single-precision floating point computations, so the Python interface to \cudaclaw{} is limited to double-precision, though it can still calculate in single-precision if desired.  Additionally, the high-performance graphics available in \cudaclaw{} are not yet available through the PyClaw interface.

\section{Performance Analysis}
\label{performance_analysis_sec}

\subsection{Roofline Model Analysis}

Analyzing the performance of algorithms on multicore machines is a subtle task, especially with the range
of different multicore architectures with their various computation and communication characteristics.
The roofline line model was proposed as an insightful performance model for multicore systems in
\cite{Williams2009Roofline}. The model gives bounds on the possible performance of a given algorithm on a
machine with given characteristics, abstracting away the specifics of the architecture. The model rates
algorithms according to their arithmetic intensity, floating point operation per byte ratio, and gives an
upper bound on the Flops/s performance. The model captures the classification of algorithms as memory-
and compute-bound where the former has low and the latter high arithmetic intensities. In this section,
we analyze the performance of \cudaclaw using the roofline model.

\begin{figure}[t]
 \begin{center}
 	\scalebox{0.44}{\includegraphics{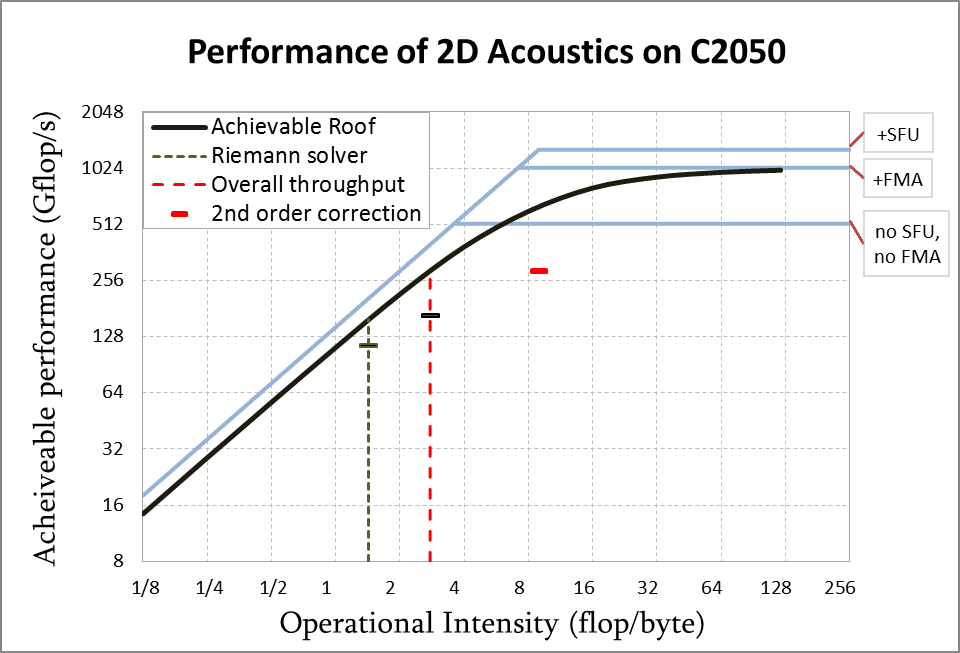}}
 \end{center}
  \caption{Achievable roofline, with achieved performance of the various components of an acoustics simulation\label{C2050_roofline_achievable_acoustics_fig}}
 \end{figure}

Our experiments are performed on the Nvidia Tesla C2050 GPU. The theoretical and achievable roofs of the C2050 are shown in figure \ref{C2050_roofline_achievable_acoustics_fig}. The achievable roof of the C2050 was recorded by using a microbenching artificial algorithm that uses solely multiply and add operations which are executed by the fused multiply add (FMA) unit of the GPU, reading and writing by coalesced and aligned accesses.


  \begin{figure}[t]
 \begin{center}
 	\scalebox{0.44}{\includegraphics{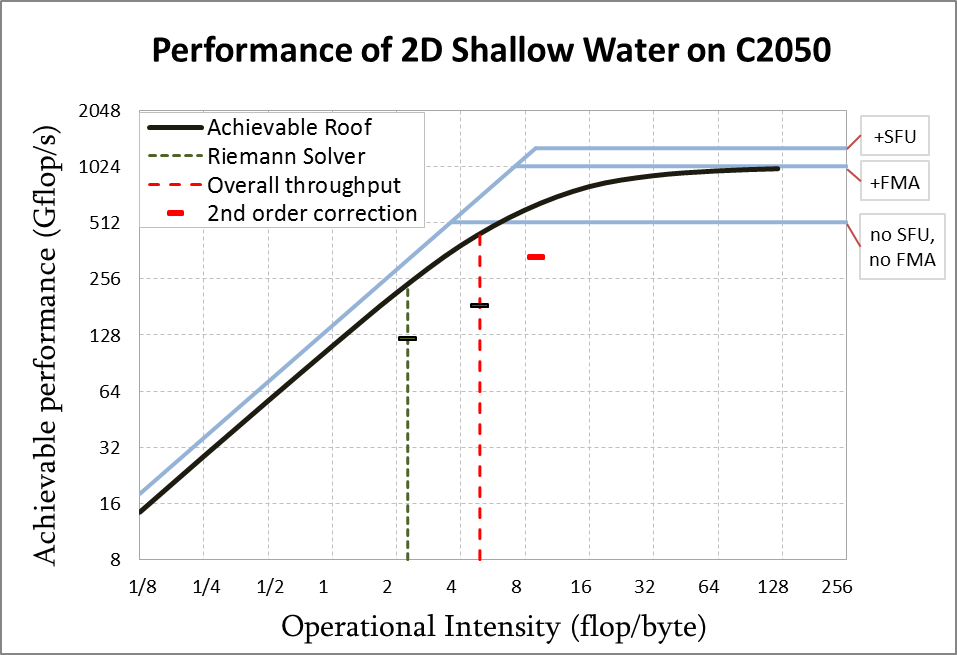}}
 \end{center}
  \caption{Achievable roofline, with achieved performance of the various components of a shallow-water  simulation\label{C2050_roofline_achievable_shallow_water_fig}}
 \end{figure}

  \begin{figure}[t]
 \begin{center}
 	\scalebox{0.44}{\includegraphics{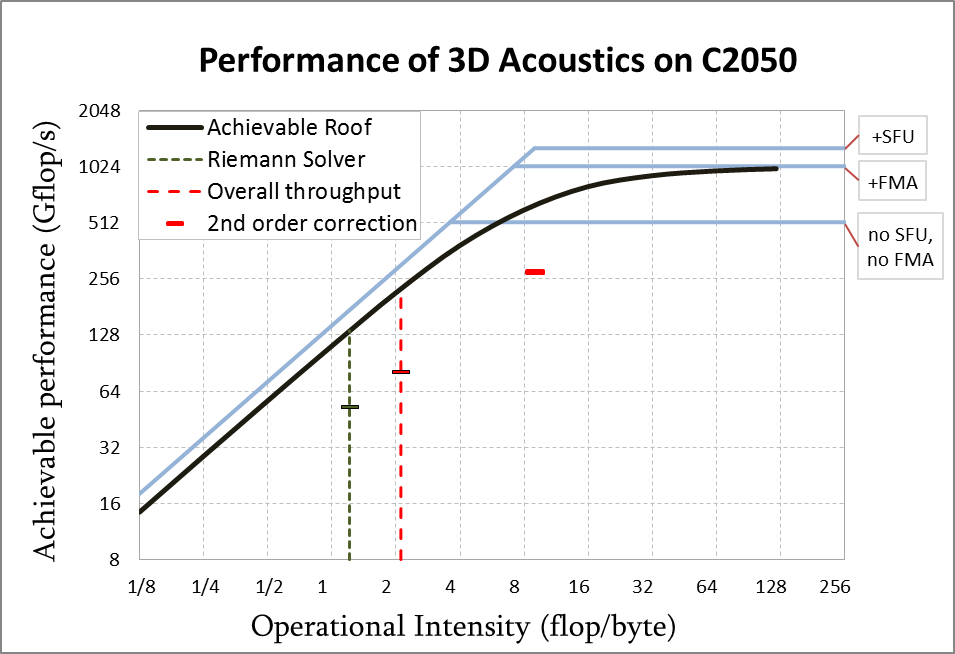}}
 \end{center}
  \caption{Achievable roofline, with achieved performance of the various components of a 3D acoustics  simulation\label{C2050_roofline_achievable_acoustics_3D_fig}}
 \end{figure}

We use the linear acoustics and shallow water flow simulations as our tests to gauge the performance of the framework. We use Nvidia's Nsight analysis tools to measure the floating point operations done, and the amount of memory read and written between the L2 cache and global device memory. Tables \ref{Acoustics_full_kernel_analysis_tbl} and \ref{Shallow_w_full_kernel_analysis_tbl} show, for each of the directional updates of the acoustics and shallow water problems respectively, the total memory size transferred, the number of floating point operations done, and the weighted average of operational intensity, for a problem size of $1024\times 1024$.

\begin{table}[!h]
\caption{2D acoustics memory (MB) and floating point operations data, problem size $=1024\times 1024$}
\label{Acoustics_full_kernel_analysis_tbl}
{
	\setlength{\extrarowheight}{1pt}
	\begin{tabular}{ |l|c|c|c| }
	\hline
	Kernel  &
	Memory & 
	MFlops &
	Operation Intensity \\
	\hline
	Horizontal & $35.1$  & $103$ &  \multirow{2}{*}{$2.77$} \\\cline{1-3}
	Vertical      & $41.4$ & $118$ & \\\hline
	\end{tabular}
}
\end{table}

\begin{table}[!h]
\caption{Shallow water memory (MB) and floating point operations data, problem size $=1024\times 1024$}
\label{Shallow_w_full_kernel_analysis_tbl}
{
	\setlength{\extrarowheight}{1pt}
	\begin{tabular}{ |l|c|c|c| }
	\hline
	Kernel  &
	Memory & 
	MFlops &
	Operation Intensity \\
	\hline
	Horizontal & $27.0$ & $153$ &  \multirow{2}{*}{$4.90$} \\\cline{1-3}
	Vertical      & $37.0$ & $175$ & \\\hline
	\end{tabular}
}
\end{table}

\begin{table}[!h]
\caption{3D acoustics memory (MB) and floating point operations data, problem size $=96\times 96\times 96$}
\label{Acoustics3D_full_kernel_analysis_tbl}
{
	\setlength{\extrarowheight}{1pt}
	\begin{tabular}{ |l|c|c|c| }
	\hline
	Kernel  &
	Memory & 
	MFlops &
	Operation Intensity \\
	\hline
	Horizontal & $34.5$ & $82$ &  \multirow{3}{*}{$2.11$} \\\cline{1-3}
	Vertical      & $40.5$ & $86$ & \\\cline{1-3}
	Depth      & $42.5$ & $86$ & \\\hline
	\end{tabular}
}
\end{table}

The operational intensity numbers in tables \ref{Acoustics_full_kernel_analysis_tbl} and \ref{Shallow_w_full_kernel_analysis_tbl} give only an average count of operations over the whole simulation. \cudaclaw's main solver kernels are composed of several stages: data reading, Riemann solution, limiter computation, first and second order flux computations, local reduction, shared memory data passing and update, each with very different Flop-to-byte ratio. From these only the Riemann solver and update stages read from and write to global memory, respectively. The other stages largely use register and shared memory as shown in \ref{solver_mem_interaction_fig}.

To better assess the framework's performance we isolate the parts which deal with global memory from the parts that don't, and measure their performance separately.  The operational intensities of the Riemann solver portions of the core kernels with the corresponding state update are shown in tables \ref{Acoustics_stripped_kernel_analysis_tbl} and \ref{Shallow_w_stripped_kernel_analysis_tbl} for 2D acoustics and shallow water simulations and in table \ref{Acoustics3D_stripped_kernel_analysis_tbl} for 3D acoustics simulation. We situate the Riemann solvers, the second-order limiting and corrections, and the overall solver in the figures \ref{C2050_roofline_achievable_acoustics_fig}, \ref{C2050_roofline_achievable_shallow_water_fig}, and \ref{C2050_roofline_achievable_acoustics_3D_fig} where the performance of the kernels are shown against the achievable roof of the Tesla C2050. The second order corrections achieve higher performance as they are primarily compute-bound and limited only by their floating point operations mix, while the Riemann solvers are memory-bound in all shown problems.

\begin{table}[!h]
\caption{Riemann solver acoustics horizontal and vertical kernels' memory (MB) and floating point operation data}
\label{Acoustics_stripped_kernel_analysis_tbl}
{
	\setlength{\extrarowheight}{1pt}
	\begin{tabular}{ |l|c|c|c| }
	\hline
	Kernel  &
	Memory & 
	MFlops &
	Operation Intensity \\
	\hline
	Horizontal & $35.0$  & $52$ &  \multirow{2}{*}{$1.425$} \\\cline{1-3}
	Vertical      & $41.3$ & $62$ & \\\hline
	\end{tabular}
}
\end{table}

\begin{table}[!h]
\caption{Riemann solver for shallow water horizontal and vertical kernels' memory (MB) and floating point operation data}
\label{Shallow_w_stripped_kernel_analysis_tbl}
{
	\setlength{\extrarowheight}{1pt}
	\begin{tabular}{ |l|c|c|c| }
	\hline
	Kernel  &
	Memory & 
	MFlops &
	Operation Intensity \\
	\hline
	Horizontal & $27.2$  & $76$ &  \multirow{2}{*}{$2.49$} \\\cline{1-3}
	Vertical      & $36.8$ & $91$ & \\\hline
	\end{tabular}
}
\end{table}

\begin{table}[!h]
\caption{Riemann solver for 3D acoustics horizontal, vertical and depth kernels' memory (MB) and floating point operation data}
\label{Acoustics3D_stripped_kernel_analysis_tbl}
{
	\setlength{\extrarowheight}{1pt}
	\begin{tabular}{ |l|c|c|c| }
	\hline
	Kernel  &
	Memory & 
	MFlops &
	Operation Intensity \\
	\hline
	Horizontal & $34.5$ & $49$ &  \multirow{3}{*}{$1.21$} \\\cline{1-3}
	Vertical      & $40.5$ & $52$ & \\\cline{1-3}
	Depth      & $42.5$ & $44$ & \\\hline
	\end{tabular}
}
\end{table}
As can be seen from the plots, the various portions of \cudaclaw{} achieve very respectable performance---near the peak performance achievable for their arithmetic intensity.  

The observed performance gap between \cudaclaw{} and the achievable performance can be attributed to a few factors. First, the achieved performance was obtained by running an artificial algorithm with only addition and multiplication operation which are executed together by the FMA unit of the processors. This algorithm does not reflect the floating point composition of our kernels, of which about $51\%$ are simple addition or multiplication operation, and $14 \%$ are special functions (square root, division,\ldots). Although special functions are executed by the SFU units of the GPU and can in theory run simultaneously with the other compute units, in practice data dependency restricts them to wait or stall the execution, in fact another test showed us that even with no data dependency, the maximum throughput of special function did not exceed $97 Gflop/s$, hence, any kernel with a significant portion of such functions will be severely limited by their low throughput. Second, all stages involve large numbers of integer operations required for addressing, amounting to up to $60\%$ (requires clear counting protocol) of all operations. Also the stages of the kernels involve inter-stage and intra-stage (local reduction) synchronizations, increasing the kernels' execution time. Third, the access pattern we have for a stencil based computation of this nature will have to sacrifice one dimension, in our case the horizontal direction, with cache line sizes, making less than optimal memory bandwidth usage. This is shown by the number of transactions required for the horizontal kernel, which requires almost 2 memory transaction per request, whereas the vertical kernels do a perfect one-to-one transaction per request.

\subsection{Numerical Experiments}
For our experiments we use two CUDA graphics cards: the Tesla C2050 and the GTX 460. These cards cover the mid to high end and low to mid ranges respectively. Tables \ref{exp_system1} and \ref{exp_system2} summarize the systems on which we ran our experiments.

\begin{table}[!h]
\caption{Experimental system 1}
\label{exp_system1}
\begin{center}
\begin{tabular}{ |l|l|l| }
\hline
Component & Our System & Notes \\
\hline
CPU  & Intel i7 950 3.07 GHz & - \\
GPU  & Nvidia Tesla C2050 (3GB)& ECC Off\\
RAM  & 4GB System RAM & - \\
OS   & Windows 7 & -\\
CUDA & CUDA 5.0 & - \\
Platform & Microsoft VS 2010 & - \\
Debugger & Nvidia Nsight 2.2& - \\
\hline
\end{tabular}
\end{center}
\end{table}

\begin{table}[!h]
\caption{Experimental system 2}
\label{exp_system2}
\begin{center}
\begin{tabular}{ |l|l|l| }
\hline
Component & Our System & Notes \\
\hline
CPU  & Intel i7 920 2.66 GHz & - \\
GPU 0 & Nvidia GTX 460 (1GB)& Display\\
GPU 1 & Nvidia GTX 460 (1GB)& Computation\\
RAM  & 6GB System RAM & - \\
OS   & Windows 7 & - \\
CUDA & CUDA 4.2 & - \\
Platform & Microsoft VS 2010 & - \\
Debugger & Nvidia Nsight 2.2& - \\
\hline
\end{tabular}
\end{center}
\end{table}

\begin{figure}[t]
 \begin{center}
 	\scalebox{0.5}{\includegraphics{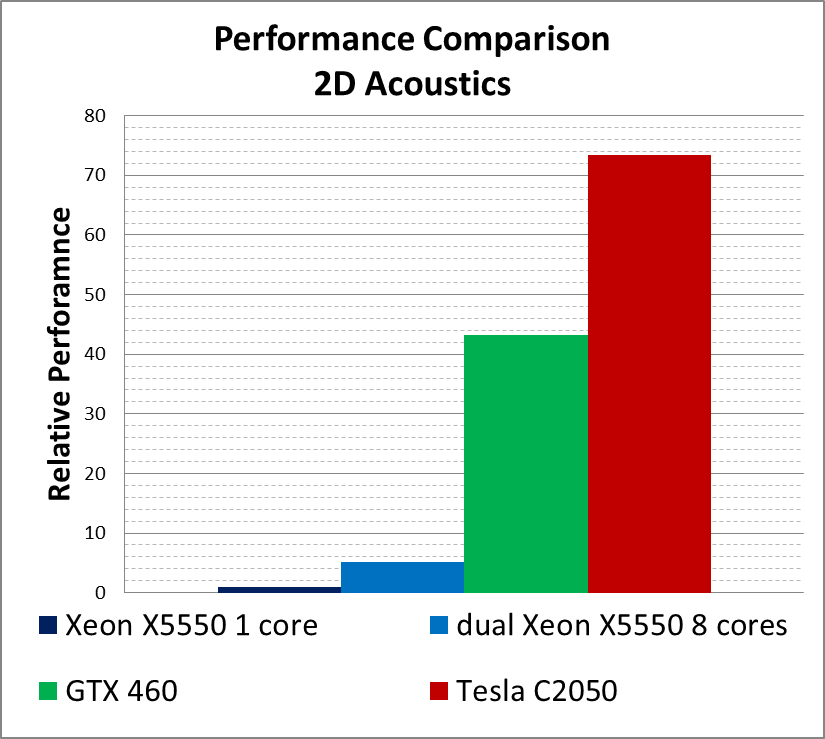}}
 \end{center}
  \caption{Double precision comparison of an acoustics simulation between a dual quad Core Xeon X5550 running PyClaw, a GTX 460 and a Tesla C2050 running CUDACLAW.\label{acoustics_rel_perf_fig} Single-precision computations on the GPUs give a substantial speedup as described in the text.  }
 \end{figure}

\begin{figure}[t]
  \begin{center}
  	\scalebox{0.5}{\includegraphics{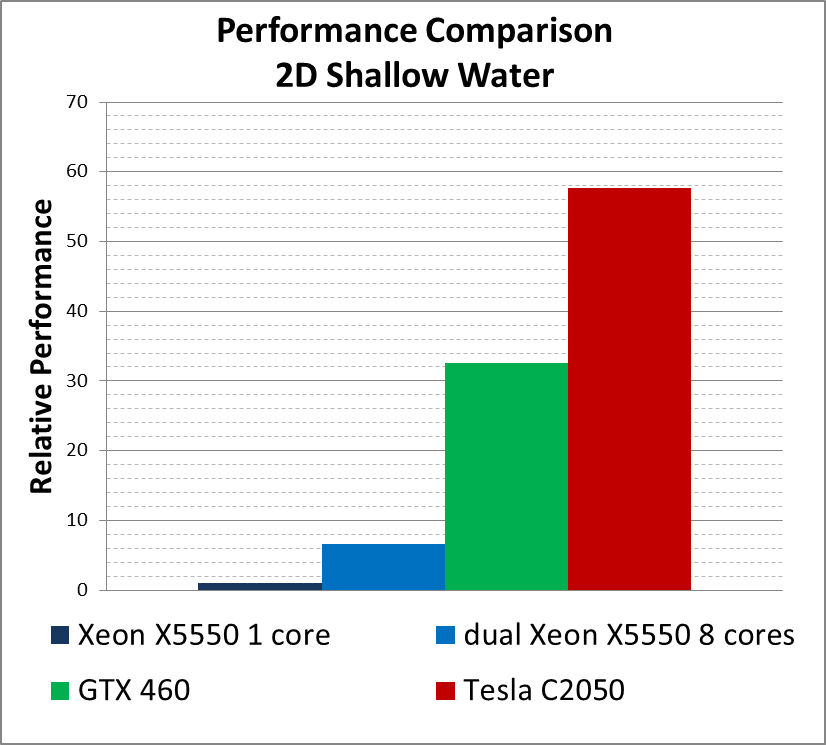}}
  \end{center}
   \caption{Double precision comparison of a shallow water simulation between a dual quad Core Xeon X5550 running PyClaw, a GTX 460 and a Tesla C2050 running CUDACLAW. \label{shallow_water_rel_perf_fig} Single-precision computations on the GPUs give a substantial speedup as described in the text.}
  \end{figure}

\begin{figure}[t]
 \begin{center}
 	\scalebox{0.45}{\includegraphics{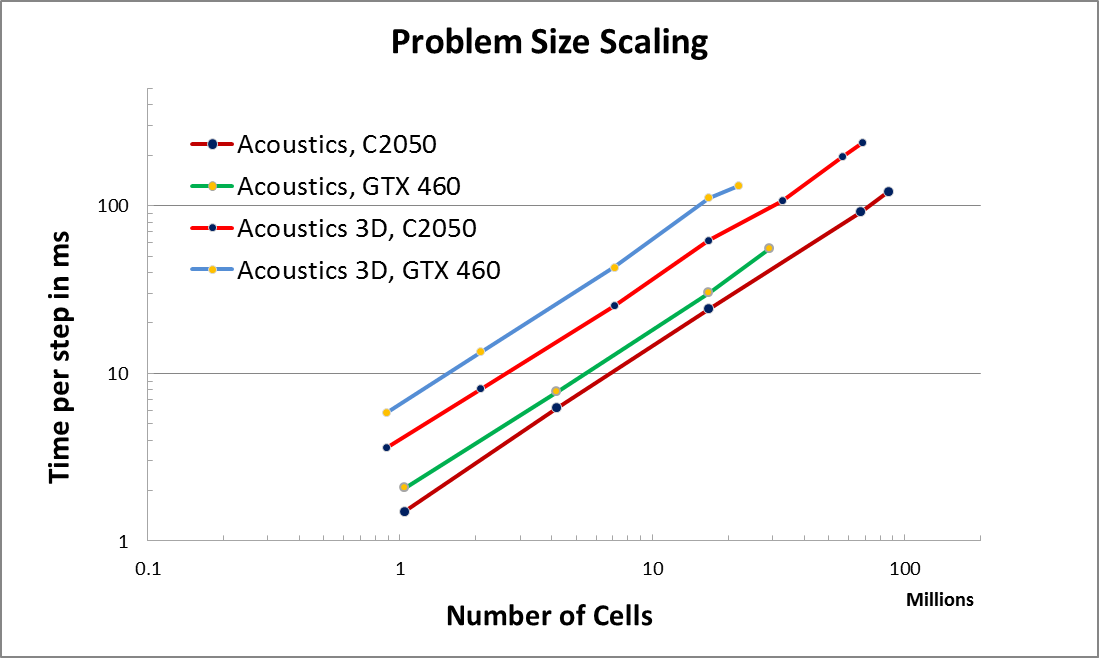}}
 \end{center}
  \caption{Problem size scaling achieved by CUDACLAW for 2D/3D acoustics simulations\label{acoustics_size_scaling_fig}}
 \end{figure}

\begin{figure}[t]
 \begin{center}
 	\scalebox{0.45}{\includegraphics{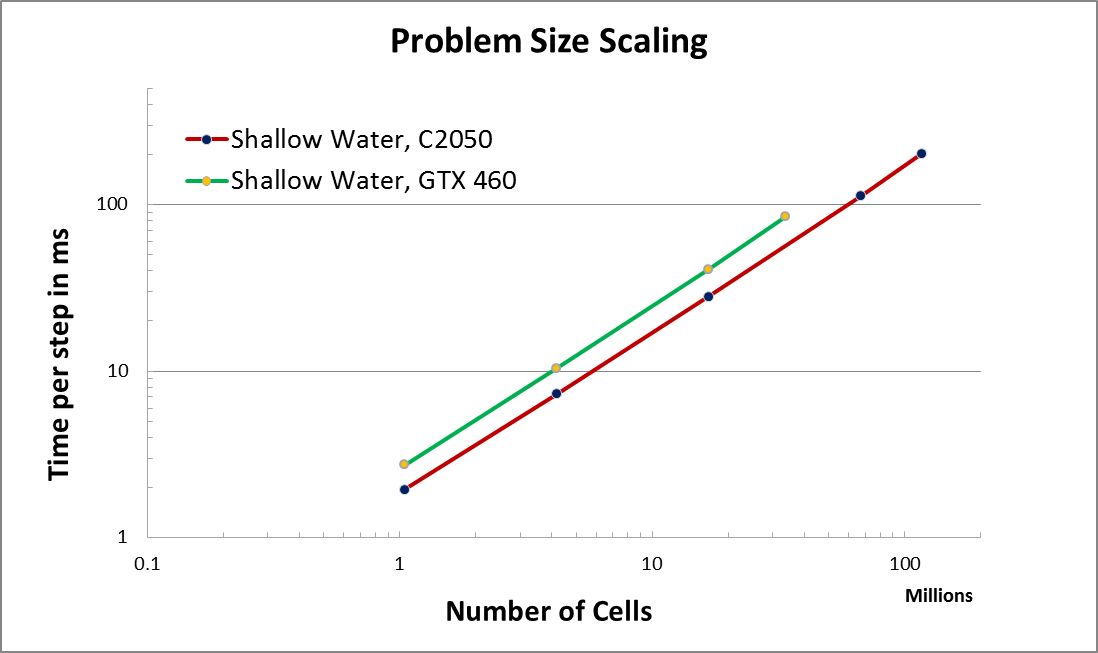}}
 \end{center}
  \caption{Problem size scaling achieved by CUDACLAW for the shallow water simulation\label{shallow_water_size_scaling_fig}}
 \end{figure}

We measure the average time step to solve an acoustics and shallow water simulations of size $1024\times 1024$ on a CPU and both our GPU, both in single and double precision. Comparative double precision results are shown in figures \ref{acoustics_rel_perf_fig} and \ref{shallow_water_rel_perf_fig}. The double precision performance of the GTX 460 is lacking because it has a Fermi 2.1 SM with low double precision throughput. Single precision throughput of the GPUs is substantially higher, with the Tesla C2050 performing at 158 and 182 GFlop/s on the 2D acoustics and shallow water simulations respectively. In figures \ref{acoustics_size_scaling_fig} and \ref{shallow_water_size_scaling_fig}, we show how the GPU handles problems of increasing size, with some superlinear scaling behaviour due to a better work distribution and latency hiding with the larger sizes. 

We compare our numbers to the results obtained in \cite{Rostrup2010}, where the shallow water equations were solved using a manually tuned code on the Tesla C2050. On a $1000\times1000$ grid, the average time step takes 3.32 ms in single precision and 8.10 ms in double precision. In comparison, \cudaclaw{} takes an average time step of 1.9 ms in single precision and 9.2 ms in double precision, showing how \cudaclaw{} performs on par with manually tuned code. In addition, the reduced memory footprint of \cudaclaw{} is seen by the fact that it can run these shallow water simulations on grids of size greater than $10,000\times 10,000$ (Figure \ref{shallow_water_size_scaling_fig}, while the simulation in \cite{Rostrup2010} can only reach grids smaller than $5,000\times 5,000$.

\section{Future Work}
\label{future_work_conclusion_sec}

We have plans to extend this work in a number of directions. One is the automatic generation of the Riemann solvers from a mathematical description of the PDEs, following the example of FEnICS~\cite{fenics} for elliptic PDEs. Even tough, we have eliminated the need to use CUDA constructs in writing Riemann solvers in \cudaclaw{}, serial procedural code is needed when solving a new set of PDEs.  A module to generate this procedural code on demand from a declarative specification of the PDEs will likely make \cudaclaw{} accessible to a much broader set of users. Similarly, boundary conditions, now limited to those provided by \cudaclaw, should be specifiable by algebraic equations. 

Adaptive spatial refinement is an area of significant practical importance. Our framework is currently limited by the initial
grid size at launch which could potentially be insufficient for certain problems that require high resolutions in localized
regions such as shock wave fronts. Using a high fixed resolution everywhere is wasteful in terms of memory usage and is
generally impractical. The solution is to use adaptive mesh refinement (AMR) where we detect regions where higher resolutions
would be needed, and refine the mesh in those regions. The adaptive nature of AMR requires dynamic memory structure creation
and manipulation. With the advent of the new Kepler architecture and CUDA 5, kernel launch through other kernels has
become possible and adaptive mesh refinement can make direct use of this new feature.

Finally, multi-GPU implementations are needed to enable \cudaclaw{} on modern supercomputers that are integrating GPU
coprocessors in their nodes. Using multiple GPUs adds a layer of complications in dealing with data splitting, communication
and balancing loads. We project from our current experience that the best way to implement such a distributed computation is to
give each node an independent subproblem for some number of time steps. Although the latency in inter-node communication gets higher, multiple local time steps can keep the nodes busy for a longer period of time and with a careful balance can make
computation and communication almost completely overlap.

\section*{Acknowledgements.} We would like to thank Rio Yokota for insights into the roofline model, Jeff Stuart for an early adaptation of Clawpack on the GPU, Wajih Bou Karam for his help with graphics inter-operability, and Andy Terrel for useful discussions on manycore architectures.

\addcontentsline{toc}{chapter}{References}

\bibliographystyle{plain}
\bibliography{CUDACLAW_paper}

\end{document}